# Multi-Cohort Development and Validation of 2D and 3D Deep Learning Models for MRI-based Parkinson's Disease Classification: Comparative Analysis of ConvKANs, CNNs, and GCNs

SB Patel[1,3], V Goh[2,3], JJ FitzGerald[1,4], CA Antoniades[1*]

Word count (excluding title page, abstract, methods, references, figures and tables): 3,141

[*] Corresponding author: Prof Chrystalina Antoniades, Nuffield Department of Clinical Neuroscience, University of Oxford, Oxford, United Kingdom. Tel: 01865 234829 Email: chrystalina.antoniades@ndcn.ox.ac.uk

[1] NeuroMetrology Lab, Nuffield Department of Clinical Neurosciences, University of Oxford, Oxford, United Kingdom

[2] School of Biomedical Engineering and Imaging Sciences, King's College London, King's Health Partners, Lambeth Wing, St Thomas' Hospital, London, SE1 7EH UK

[3] Department of Radiology, Guy's and St Thomas' NHS Foundation Trust, Lambeth Wing, St Thomas' Hospital, London, SE1 7EH UK

[4] Nuffield Department of Surgical Sciences, University of Oxford, Oxford, United Kingdom

# Abstract

Parkinson's Disease (PD) diagnosis remains challenging. This study applies Convolutional Kolmogorov-Arnold Networks (ConvKANs), integrating learnable spline-based activation functions into convolutional layers, for PD classification using structural MRI. The first 3D implementation of ConvKANs for medical imaging is presented, comparing their performance to Convolutional Neural Networks (CNNs) and Graph Convolutional Networks (GCNs) across three open-source datasets. Isolated analyses assessed performance within individual datasets, using cross-validation techniques. Holdout analyses evaluated cross-dataset generalizability by training models on two datasets and testing on the third, mirroring real-world clinical scenarios. In isolated analyses, 2D ConvKANs achieved the highest AUC of 0.99 (95% CI: 0.98-0.99) on the PPMI dataset, outperforming 2D CNNs (AUC: 0.97, p = 0.0092). 3D models showed promise, with 3D CNN and 3D ConvKAN reaching an AUC of 0.85 on PPMI. In holdout analyses, 3D ConvKAN demonstrated superior generalization, achieving an AUC of 0.85 on early-stage PD data. GCNs underperformed in 2D but improved in 3D implementations. These findings highlight ConvKANs' potential for PD detection, emphasize the importance of 3D analysis in capturing subtle brain changes, and underscore cross-dataset generalization challenges. This study advances AI-assisted PD diagnosis using structural MRI and emphasizes the need for larger-scale validation.

# Introduction

Parkinson's Disease (PD) is the second most common neurodegenerative disorder, affecting over 10 million people worldwide.[1] Clinically, PD is characterized by motor symptoms such as tremor, rigidity, and bradykinesia, as well as non-motor symptoms including cognitive impairment and depression.[2] The prevalence of PD increases with age, and as the global population ages, the burden of PD is expected to increase substantially.

Early and accurate diagnosis of PD remains challenging, as current diagnostic criteria rely on subjective clinical assessment of motor symptoms, which often emerge only after significant neurodegeneration has occurred.[3] Misdiagnosis rates can be as high as 25% in early stages, highlighting the need for objective biomarkers to support clinical decision-making.[4] While various imaging modalities have been explored, structural Magnetic Resonance Imaging (MRI) is not part of any PD diagnostic criteria due to the subtle and heterogeneous nature of early brain changes.[5]

Deep learning has emerged as a powerful methodology for medical image analysis, demonstrating success across various domains including neuroimaging.[6,7] Convolutional Neural Networks (CNNs) have been widely applied to MRI analysis, using their ability to capture hierarchical features.[8] CNNs' rigid structure may limit their ability to model non-linear relationships in high-dimensional data.[9]

Graph-based approaches have recently gained traction in the deep imaging community, with Graph Convolutional Networks (GCNs) offering a framework for modeling the inherent structural relationships in medical imaging data, including in PD.[10] By representing an image or scan as a graph of interconnected nodes, GCNs can capture both local and global context, potentially overcoming some limitations of traditional CNN architectures.[11] However, the application of GCNs in neuroimaging is still an emerging field, with limited studies exploring their potential for PD classification.

The Kolmogorov-Arnold Network (KAN), introduced earlier this year, represents a significant departure from traditional CNN architectures.[12] Based on the Kolmogorov-Arnold representation theorem, KANs replace conventional weight matrices with learnable spline functions, offering enhanced flexibility in modeling complex, non-linear relationships.

Building upon the KAN framework, the Convolutional Kolmogorov-Arnold Network (ConvKAN) was recently proposed as a fusion of KAN principles with convolutional architectures.[13] ConvKANs integrate spline-based functions into convolutional layers, combining the flexibility of KANs with the spatial invariance of CNNs. While ConvKANs have shown promising results in 2D image analysis tasks, their application to 3D imaging data, such as volumetric MRI scans, represents a novel and unexplored step.[14]

The dimensionality of input data is another critical consideration in imaging analysis. Studies comparing 2D and 3D approaches have yielded mixed results, with some favoring slice-based methods for their computational efficiency and larger effective sample sizes, while others advocate for 3D analysis to capture spatial relationships and avoid information loss.[15,16] The relative performance of different architectures across 2D and 3D implementations also remains unclear, with few studies conducting comprehensive comparisons.[17]

To address these knowledge gaps, we present a comprehensive evaluation of deep learning architectures for MRI-based PD classification, with a focus on the novel application of ConvKANs. We compare the performance of ConvKANs, CNNs, and GCNs across both 2D and 3D implementations, using multiple open-source datasets to assess within-dataset performance and cross-dataset generalizability. Furthermore, we introduce the first 3D implementation of ConvKANs, exploring their potential for volumetric MRI analysis.

By conducting this multi-cohort, comparative study, we aim to provide valuable insights into the optimal approach for deep learning-based PD diagnosis using structural MRI. The identification of robust, generalizable models could pave the way for AI-assisted diagnostic tools, supporting early detection and intervention in PD.

# Results

## Dataset Characteristics

This study utilised three open-source datasets: the Parkinson's Progression Markers Initiative (PPMI) MRI dataset, NEUROCON and Tao Wu (Table 1).[18,19] While all three datasets included both PD patients and age-matched healthy control subjects, there were differences in patient characteristics. The PPMI cohort was restricted to newly diagnosed PD patients within 2 years of diagnosis who had not yet started any PD medications. The NEUROCON and Tao Wu datasets included PD patients with longer average disease durations, most of whom were already receiving treatment with dopaminergic medications such as levodopa.

Table 1: *Participant demographics in each dataset.*

We compared 2D and 3D implementations of Convolutional Neural Networks (CNN), Convolutional Kolmogorov-Arnold Networks (ConvKAN), and Graph Convolutional Networks (GCN). For 2D analyses, our large sample sizes (PPMI: n=5900, Tao Wu: n=4000, NEUROCON: n=4300) provided statistical power >0.99 to detect even small effect sizes (Cohen's d = 0.50). 3D analyses, constrained by computational resources, used smaller samples (PPMI: n=59, Tao Wu: n=40, NEUROCON: n=43), achieving adequate power (0.69-0.85) for large effect sizes (Cohen's d = 0.80), but limited power for smaller effects.

# Isolated Dataset Analysis

## PPMI Dataset Performance

In the PPMI dataset, we observed performance differences across models and dimensionalities. The 2D ConvKAN achieved the highest accuracy of 0.93 (95% CI: 0.92-0.94) and AUC of 0.99 (95% CI: 0.98-0.99). This performance significantly surpassed the 2D CNN, which achieved an accuracy of 0.90 ($p = 0.0075$, Cohen's d = 1.3) and AUC of 0.97 ($p = 0.0092$, Cohen's d = 1.1). The 2D GCN models showed lower performance, with an accuracy of 0.53 and AUC of 0.54, significantly underperforming compared to both 2D ConvKAN ($p < 0.01$, Cohen's d = 12 for accuracy) and 2D CNN ($p < 0.01$, Cohen's d = 12 for accuracy).

Comparing 2D and 3D implementations, the 2D ConvKAN significantly outperformed its 3D counterpart in accuracy (0.93 vs 0.85, $p = 0.045$, Cohen's d = 2.0) and recall (0.93 vs 0.85, $p = 0.045$, Cohen's d = 2.0). For AUC, the difference was not statistically significant (0.99 vs 0.85, $p = 0.81$, Cohen's d = 0.63).

Among 3D models, the CNN achieved the highest accuracy of 0.85, while the ConvKAN attained the highest AUC of 0.85. The differences between 3D models were not statistically significant ($p > 0.05$ for all comparisons).

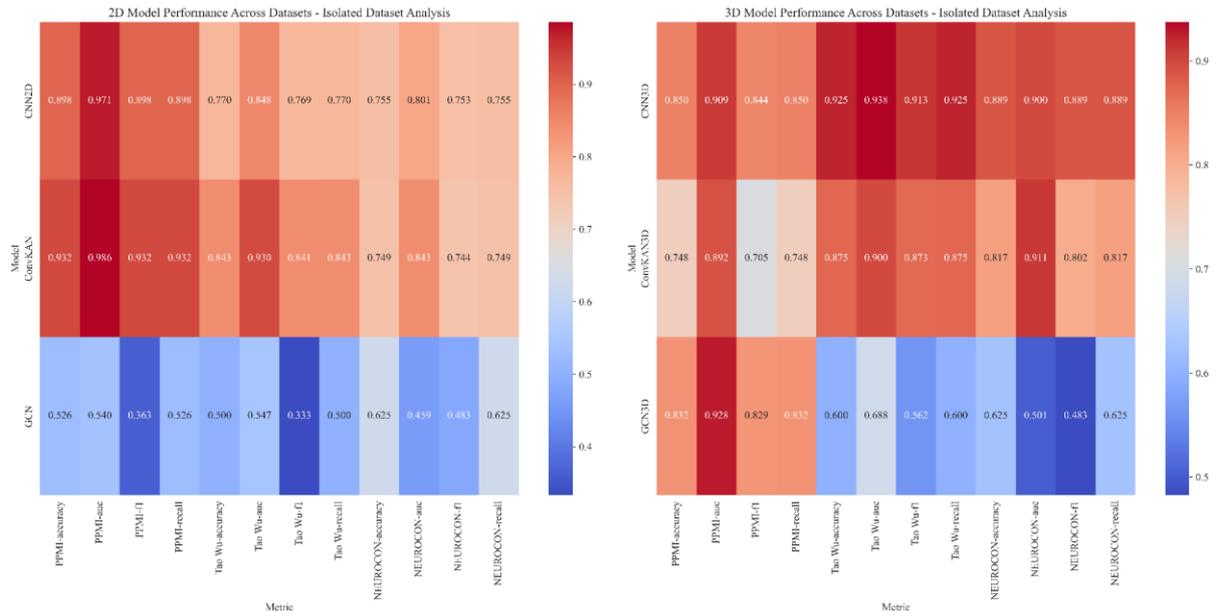

Figure 1: *Comparison of 2D and 3D model performance across datasets in isolated dataset analysis. Heatmap shows accuracy, F1 score, recall, and AUC for each model type across the PPMI, Tao Wu, and NEUROCON datasets.*

## Tao Wu Dataset Performance

In the Tao Wu dataset, the 2D ConvKAN maintained strong performance with an accuracy of 0.84 and AUC of 0.93. The difference in performance compared to 2D CNN (accuracy: 0.84; AUC: 0.90) was statistically significant for AUC ($p = 0.034$, Cohen's d = 1.1) but not for accuracy ($p = 0.21$, Cohen's d = 0.68). The 2D GCN showed significantly lower performance (accuracy: 0.50; AUC: 0.46) compared to both ConvKAN and CNN ($p < 0.01$ for all comparisons).

3D models generally outperformed their 2D counterparts, though not all differences were statistically significant. The 3D CNN achieved the highest accuracy of 0.93 and an AUC of 0.94, surpassing its 2D version (accuracy: 0.93 vs. 0.84, $p = 0.19$, Cohen's d = 1.1; AUC: 0.94 vs. 0.90, $p = 0.19$, Cohen's d = 0.76). The 3D ConvKAN showed similar performance to its 2D counterpart (accuracy: 0.88 vs. 0.84, $p = 0.63$, Cohen's d = 0.15). The 3D GCN demonstrated improvement over the 2D GCN (accuracy: 0.77 vs. 0.50, $p = 0.37$, Cohen's d = 0.63).

Among 3D models, while the CNN showed the highest performance, the differences were not statistically significant ($p > 0.05$ for all comparisons).

## NEUROCON Dataset Performance

In the NEUROCON dataset, we observed less pronounced differences. The 2D ConvKAN showed higher AUC (0.84) compared to 2D CNN (0.80), although this difference was not statistically significant ($p = 0.31$, Cohen's $d = 0.27$). 2D ConvKAN outperformed 2D GCN in both accuracy (0.75 vs 0.63, $p < 0.01$, Cohen's $d = 1.5$) and AUC (0.84 vs 0.55, $p < 0.01$, Cohen's $d = 4.3$).

3D models generally outperformed their 2D counterparts, but with less pronounced differences. The 3D CNN achieved the highest accuracy of 0.89, outperforming its 2D version (0.89 vs 0.75, $p = 0.31$, Cohen's $d = 0.69$). The 3D ConvKAN showed similar performance improvements over its 2D counterpart (accuracy: 0.86 vs 0.75, $p = 0.63$, Cohen's $d = 0.31$), while the 3D GCN demonstrated higher accuracy compared to its 2D version (0.83 vs 0.63, p-value not available due to identical values).

Among 3D models, while the CNN achieved the highest accuracy (0.89) and the ConvKAN attained the highest AUC (0.91), the differences were not statistically significant ($p > 0.05$ for all comparisons).

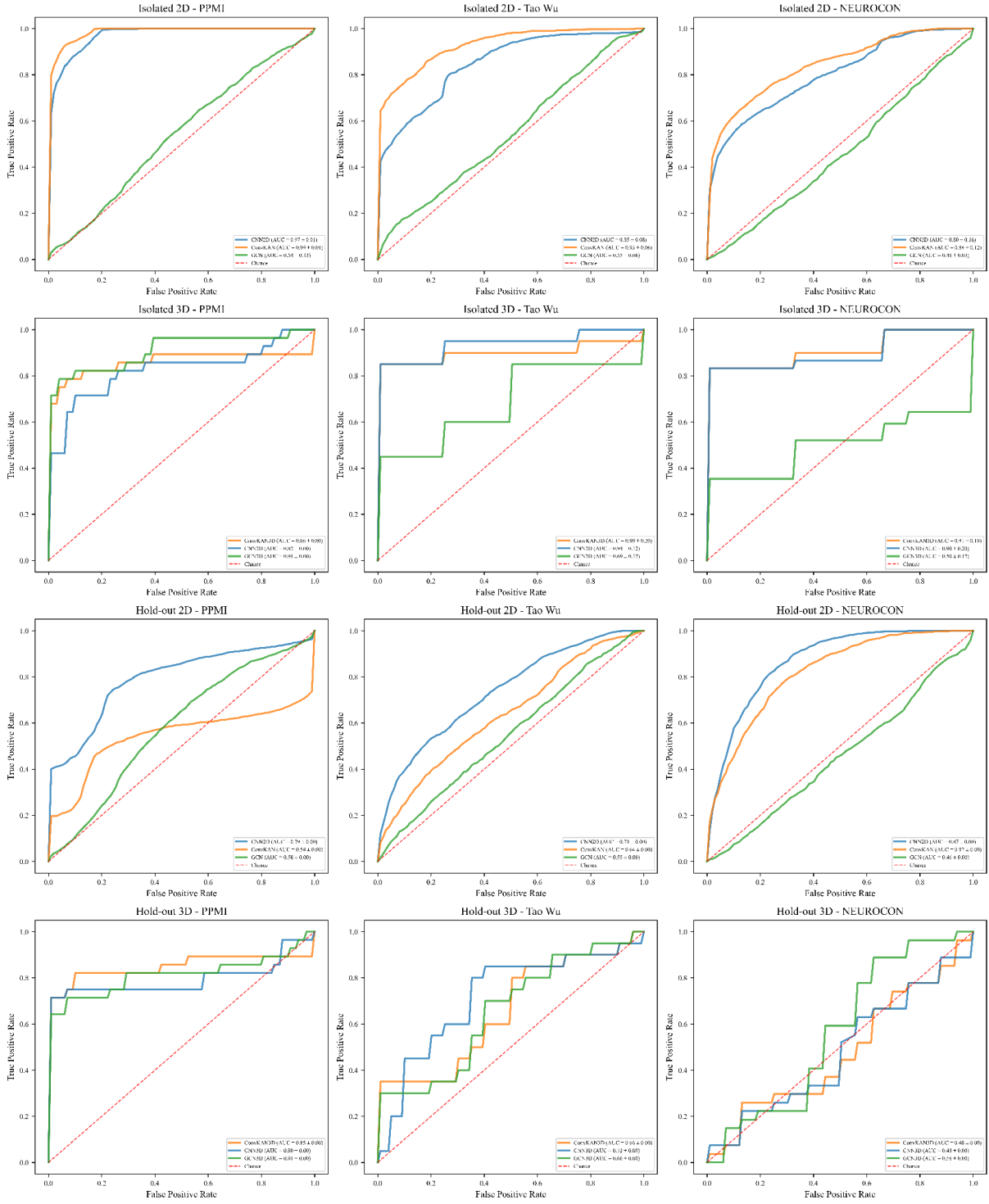

Figure 2: *Receiver Operating Characteristic (ROC) curves for 2D and 3D models across all datasets in isolated and hold-out analyses. Each subplot shows the true positive rate against the false positive rate at various threshold settings for CNN (blue), ConvKAN (orange), and GCN (green) models.*

# Hold-Out Analysis

In the hold-out analysis, where models were trained on two datasets and tested on the third, we observed a general decrease in performance but with some notable exceptions. It's important to note that due to the nature of this analysis, we had limited data points (one per model per test dataset), which precluded the use of statistical tests for comparing model performances. This limitation arises from the absence of multiple folds or repeated measurements in the hold-out scenario, unlike in the isolated dataset analysis where we had multiple folds for each dataset.

When tested on the PPMI dataset, the 3D ConvKAN demonstrated superior generalization capabilities, achieving the highest accuracy of 0.83 and AUC of 0.85. This performance notably outperformed its 2D counterpart (accuracy: 0.59, AUC: 0.54). Among 2D models, the CNN showed the most consistent performance, achieving an accuracy of 0.70 and AUC of 0.79, outperforming both 2D ConvKAN and 2D GCN (accuracy: 0.47, AUC: 0.51).

When tested on the Tao Wu dataset, all models showed decreased performance compared to the isolated analysis. The 2D ConvKAN achieved the highest accuracy of 0.57, while the 2D CNN attained the highest AUC of 0.74. The 3D models showed varied performance, with 3D ConvKAN achieving the highest accuracy (0.65) among all models.

For the NEUROCON dataset, the 2D CNN demonstrated the best performance in the hold-out analysis, with an AUC of 0.87. The 2D GCN achieved the highest accuracy (0.63) among 2D models. The 3D models showed comparable performance to their 2D counterparts, with the 3D ConvKAN achieving an accuracy of 0.63 and AUC of 0.56.

These results highlight the varying generalization capabilities of different models across datasets, with 3D models showing particular promise in some scenarios, especially when generalizing to the PPMI dataset, which consists of early-stage PD patients.

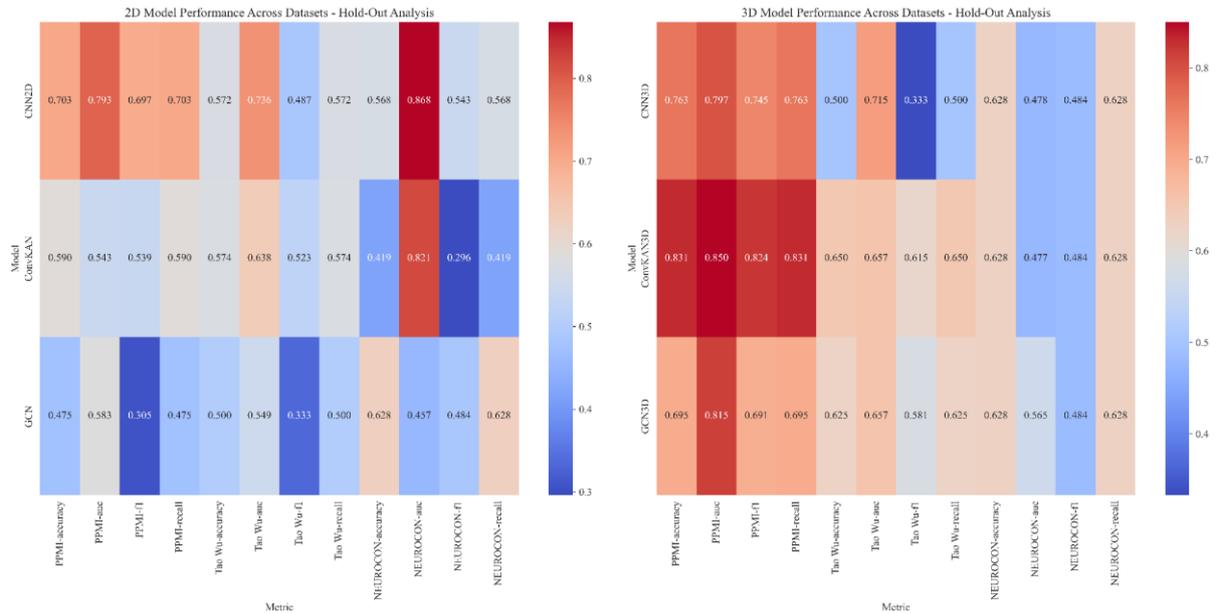

Figure 3: *Heatmaps depicting model performance across datasets in the hold-out analysis. The heatmaps display key metrics, including accuracy, F1 score, recall, and AUC, for each model type when trained on two datasets and tested on the third. Color intensity corresponds to the magnitude of each metric, with darker shades indicating higher performance values. These visualizations allow for a clear comparison of model generalizability across different datasets.*

## Model Consistency Analysis

To assess model consistency across datasets, we calculated the coefficient of variation (CV) for each model's performance in both isolated dataset and holdout analyses. CV, the ratio of the standard deviation to the mean, indicates stability, with lower values representing higher consistency.

In the 2D models, ConvKAN had the most consistent performance with CVs of 8.90% for accuracy and 6.40% for AUC. CNN followed with CVs of 7.94% and 8.23%, respectively, while GCN had the highest variability (9.79% accuracy, 7.74% AUC). For 3D models, CNN was the most consistent, showing CVs of 3.45% for accuracy and 1.74% for AUC. ConvKAN performed similarly with CVs of 6.36% and 0.88%, while GCN showed much higher variability at 15.15% for accuracy and 24.74% for AUC.

For 2D models, CNN showed the highest consistency with CVs of 10.22% for accuracy and 6.79% for AUC, followed by GCN (12.56% accuracy, 10.06% AUC). ConvKAN was more variable, with CVs of 14.59% and 17.32%. In 3D models, GCN was the most consistent (4.97% accuracy, 15.18% AUC), while ConvKAN (12.91%, 23.05%) and CNN (17.02%, 20.40%) were more variable.

Models had higher consistency in the isolated analysis compared to the holdout analysis, suggesting cross-dataset generalization is more challenging. In the isolated analysis, 3D CNN was the most stable, while ConvKAN showed high variability in holdout tests. GCN, although variable in isolated datasets, performed well in holdout conditions, especially in 3D. Comprehensive statistical analyses and performance metrics for all models are available in the Supplementary Material 1.

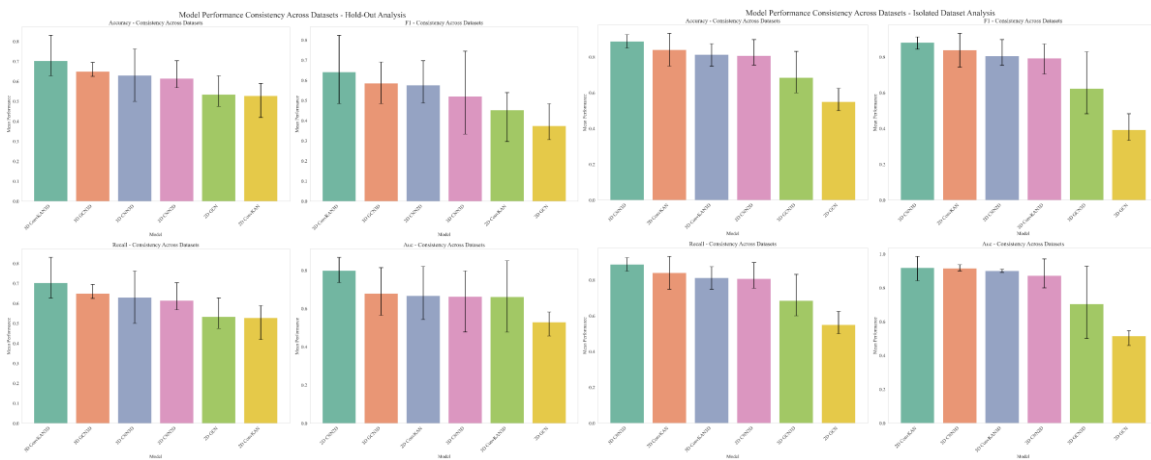

Figure 4: *Bar plots illustrating the coefficient of variation (CV) for accuracy and AUC across models and datasets in isolated and hold-out analyses. Lower CV values indicate greater model consistency. Each bar is accompanied by error bars representing the 95% confidence intervals, providing a visual representation of the models' stability and performance variability across datasets.*

## Early-Stage PD Performance

Given the unique composition of the PPMI dataset, which consists of newly diagnosed PD patients, we were able to highlight model performance in early-stage PD detection.

In the isolated dataset analysis of PPMI, the 2D ConvKAN demonstrated exceptional performance in early-stage PD detection, achieving the highest AUC of 0.99 (95% CI: 0.98-0.99) and accuracy of 0.93 (95% CI: 0.92-0.94). This performance significantly outperformed the 2D CNN, which achieved an AUC of 0.97 (p = 0.0092, Cohen's d = 1.1) and accuracy of 0.90 (p = 0.0075, Cohen's d = 1.3). The effect sizes were large for both metrics. The 2D GCN models showed substantially lower performance (AUC: 0.54, accuracy: 0.53) compared to both 2D ConvKAN and 2D CNN (p < 0.01 for all comparisons).

Among 3D models, the ConvKAN and CNN showed similar performance, with both achieving an AUC of 0.85. The differences between 3D models were not statistically significant (p > 0.05 for all comparisons).

In the hold-out analysis, where the PPMI dataset was used as the test set, the 3D ConvKAN showed the best generalization to early-stage PD, achieving an accuracy of 0.83 and AUC of 0.85. This outperformed both its 2D counterpart (accuracy: 0.59, AUC: 0.54) and other models, including the 2D CNN (accuracy: 0.70, AUC: 0.79). However, due to the limited data points in the hold-out analysis, statistical significance could not be determined for these comparisons.

These findings highlight the potential of ConvKAN models for early detection of PD using structural MRI, with 2D implementations showing superior performance in isolated dataset analysis and 3D implementations demonstrating better generalization in hold-out scenarios. The consistent trend and large effect sizes warrant further investigation. The ability to accurately identify PD in its early stages could have significant implications for clinical practice, potentially enabling earlier intervention and improved patient outcomes.

# Discussion

This comprehensive evaluation of deep learning architectures for Parkinson's Disease (PD) classification using structural MRI reveals important insights into the potential and challenges of AI-assisted diagnosis in neurodegenerative disorders. The novel application of Convolutional Kolmogorov-Arnold Networks (ConvKANs) to MRI analysis represents a significant contribution to the field, demonstrating promising performance across both 2D and 3D implementations.

In the isolated dataset analysis, 2D ConvKANs consistently outperformed traditional Convolutional Neural Networks (CNNs) and Graph Convolutional Networks (GCNs), achieving the highest accuracy (0.93) and area under the receiver operating characteristic curve (AUC) (0.99) for the PPMI dataset. This superior performance can be attributed to the unique architecture of ConvKANs, which combines the spatial invariance properties of CNNs with flexible, non-linear modeling capabilities through learnable B-spline functions. In the context of PD neuroimaging, this adaptability is particularly relevant for capturing subtle, progressive structural changes that may elude traditional CNN architectures.

The performance disparity between ConvKANs and GCNs in 2D analysis was particularly striking, with GCNs consistently underperforming across all datasets. This suggests that the conversion of 2D image data to graph representations may result in a loss of fine-grained spatial information crucial for detecting subtle PD-related changes. However, the improved performance of GCNs in 3D analyses, particularly in the PPMI dataset, indicates that graph-based approaches may be more suitable for capturing global structural relationships in volumetric data. This finding aligns with previous research by Huang et al., validating the potential of 3D graph-based approaches in neuroimaging analysis.[20]

The variability in model performance across datasets, particularly evident in the hold-out analysis, underscores the challenges in developing generalizable, dataset-agnostic foundation models for clinical neuroimaging. While 2D ConvKANs excelled in the isolated dataset analysis, their performance decreased more dramatically than CNNs in the hold-out analysis. Conversely, 3D ConvKANs showed promising generalization capabilities, particularly when tested on the PPMI dataset, which consists of early-stage PD patients. This variability can be attributed to several factors, including differences in image acquisition parameters, patient demographics, and disease severity distribution across datasets.

The superior performance of 3D ConvKANs in the hold-out analysis, especially for early-stage PD detection, is a key finding of this study. Achieving an AUC of 0.85 on the PPMI dataset demonstrates the potential of 3D ConvKANs to capture subtle structural changes associated with early-stage PD, even when trained on datasets with more advanced PD cases. This robustness in generalization suggests that 3D ConvKANs may be particularly well-suited for clinical applications where variability in imaging protocols and patient characteristics is common.

The model consistency analysis revealed interesting patterns across different architectures and dimensionalities. In isolated analyses, 3D CNN demonstrated the highest consistency, while ConvKAN showed more variability in hold-out tests. Notably, GCNs, despite their variable performance in isolated datasets, showed improved consistency in hold-out conditions, especially in 3D implementations. These findings highlight the complex interplay between model architecture, data dimensionality, and generalization capabilities, emphasizing the need for comprehensive evaluation strategies in developing AI models for clinical use.

A key limitation of our study is the relatively small sample size, particularly for 3D analyses. This limitation is reflected in the lower statistical power for detecting small effect sizes in 3D models and the inability to perform statistical tests in the hold-out analysis due to limited data points. Future studies should aim to validate these results on larger, more diverse cohorts to ensure clinical applicability across different patient populations. Additionally, the use of different cross-validation strategies for 2D and 3D analyses, while justified by the nature of the data, introduces potential bias in direct comparisons between dimensionalities.

From a clinical perspective, our findings highlight the importance of standardized imaging protocols and diverse patient cohorts in developing reliable AI-assisted diagnostic tools for PD. The variability in model performance across datasets emphasizes the need for multi-center validation studies and careful consideration of dataset characteristics in the development and evaluation of AI models.

The computational advantages of ConvKANs over traditional CNNs are particularly relevant in clinical settings. Their ability to achieve high performance levels with potentially fewer parameters, attributed to the use of learnable B-spline functions, could lead to faster training times and lower memory requirements. Such computational benefits could facilitate more efficient clinical workflows. The ability of ConvKANs to maintain high performance in both 2D and 3D implementations also offers flexibility in adapting to various clinical imaging protocols and legacy datasets.

Our study's focus on binary classification between PD patients and healthy controls, while a common approach, may oversimplify the complex spectrum of PD and fail to account for other neurodegenerative conditions such as atypical parkinsonian disorders. Future work should explore multi-class classification to better reflect the clinical reality of differential diagnosis in movement disorders.

The inherent "black box" nature of deep learning models, combined with this binary classification approach, raises the possibility that our models may be using non-PD related changes for classification. These could include global atrophy associated with normal aging or overlapping conditions present in the patient cohort, rather than PD-specific markers. Future studies should incorporate explainable AI techniques to elucidate the features driving model decisions, enhancing clinician trust and potentially uncovering novel PD biomarkers.

In conclusion, this study offers a thorough evaluation of deep learning architectures for MRI-based Parkinson's Disease classification, with the novel development and validation of ConvKANs marking a significant advancement in the field. Although MRI is not currently a primary diagnostic tool for PD, our findings demonstrate that with careful model selection and refinement, MRI analysis could become an integral part of a multimodal diagnostic approach. The superior performance of ConvKANs in isolated dataset analyses and the promising generalization of 3D ConvKANs in hold-out testing underscore the potential of this architecture in neuroimaging. However, the observed variability in performance across datasets and between isolated and hold-out analyses highlights the need for robust validation across diverse cohorts to ensure reliability in clinical applications.

As personalized medicine advances in the treatment of neurodegenerative disorders, the integration of advanced imaging techniques with clinical expertise offers significant potential to enhance patient outcomes through more accurate diagnosis, prognosis, and treatment planning in PD. Future research should address the limitations identified in this study, particularly by increasing the sample size for 3D analyses, exploring multi-class classification to better represent the full spectrum of parkinsonian disorders, and developing methods to improve the interpretability of deep learning models for use in clinical settings. The promising results of 3D ConvKANs in generalization tasks suggest that future work should focus on optimizing these 3D implementations for clinical applications, potentially leading to more robust and adaptable AI-assisted diagnostic tools for PD.

# Methods

## Dataset Description

This study utilized three open-source MRI datasets: the Parkinson's Progression Markers Initiative (PPMI), NEUROCON, and Tao Wu.[18,19] Each dataset included both Parkinson's Disease (PD) patients and age-matched healthy control subjects, with varying clinical characteristics across cohorts. The PPMI cohort consisted of newly diagnosed PD patients within two years of diagnosis, none of whom had started PD medication at the time of scanning. A subset of the PPMI dataset was randomly selected to match the sample size and MRI sequences of the other datasets for comparative purposes. Both the NEUROCON and Tao Wu datasets consisted of PD patients with longer average disease durations, most of whom were already receiving dopaminergic treatments, such as levodopa, at the time of the scans. This difference in disease stage and treatment status between the datasets was accounted for in the analysis.

## Image Preprocessing

For the 2D analysis, we extracted 100 axial slices centered on the midbrain from each T1-weighted volume. The midbrain was chosen as the centre point due to its known involvement in PD pathology.[21] Each slice was resampled to 224 x 224 pixels using bilinear interpolation. Intensity normalization was applied to scale pixel values to a [0, 1] range, and a 2D Gaussian filter ($\sigma = 1$ mm) was used for noise reduction. For the 3D analysis, whole-brain volumes were resampled to 128 x 128 x 128 voxels using trilinear interpolation, preserving the full spatial information of the brain while standardising the input dimensions for our 3D models. Similar to the 2D process, the 3D volumes underwent intensity normalization to [0, 1] and noise reduction using a 3D Gaussian filter ($\sigma = 1$ mm).

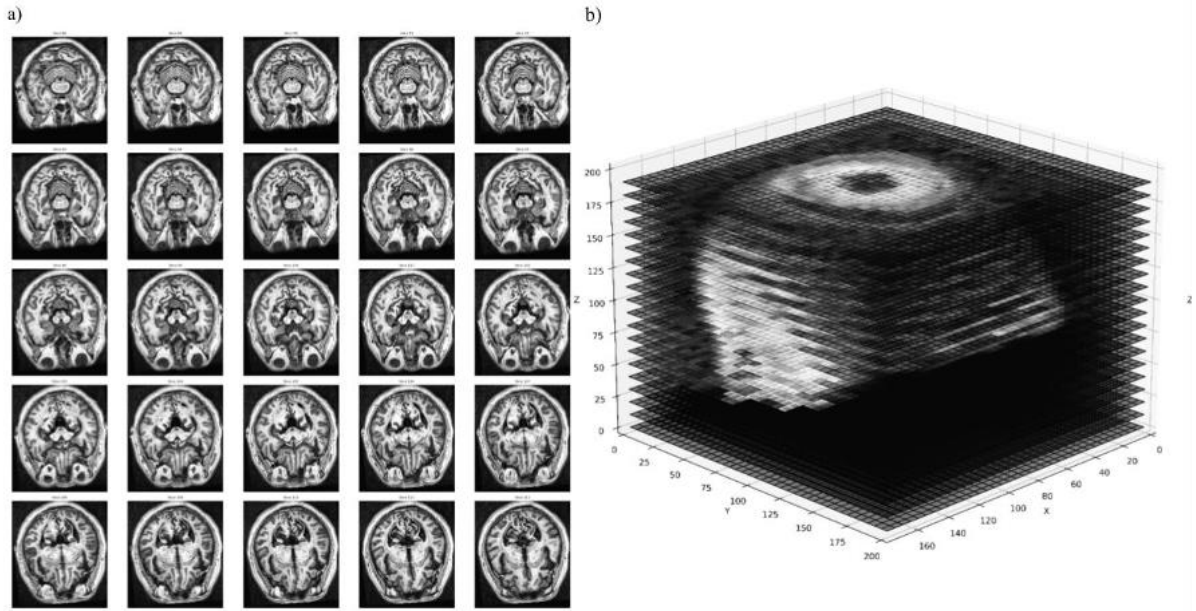

Figure 5: a) Example of single slices, centred on the midbrain, used individually to classify between PD and HC in the two-dimensional analyses. b) Entire volumetric MRI scan used in the three-dimensional analyses. Gaps are inserted at regular intervals to help visualise inner structure but are not implemented in the analyses.

## Model Architectures

We implemented Convolutional Kolmogorov-Arnold Networks (ConvKAN) in both 2D and 3D variants, extending the original ConvKAN principles to MRI analysis. Our models employ custom KANConv layers (KANConv2D for 2D and KANConv3D for 3D), which integrate traditional convolution operations with learnable B-spline functions for adaptive, non-linear transformations of input data.[22] The original design is adapted by combining the spline outputs with a SiLU (Sigmoid Linear Unit) activation, creating a hybrid mechanism that balances adaptability with stable gradient properties.

For both 2D and 3D ConvKAN implementations, we employed cubic B-spline functions (degree 3) with carefully defined knots and control points. The knots were equidistantly positioned in the range [-1, 1]. As per the standard B-spline formulation, the number of knots (n) was set to the number of control points (k) plus the degree (d) plus one, such that $n = k + d + 1$. Specifically, we used 6 control points, resulting in 10 knots for our cubic B-splines. The control points are represented by learnable weights in our model, allowing the network to adapt the shape of the spline functions during training.

The 2D ConvKAN architecture consisted of three SplineConv2d layers followed by batch normalization and max pooling. The model concluded with global average pooling and two fully connected layers. The 3D variant used four KANConv3D layers followed by batch normalization, max pooling, global average pooling, and two fully connected layers.

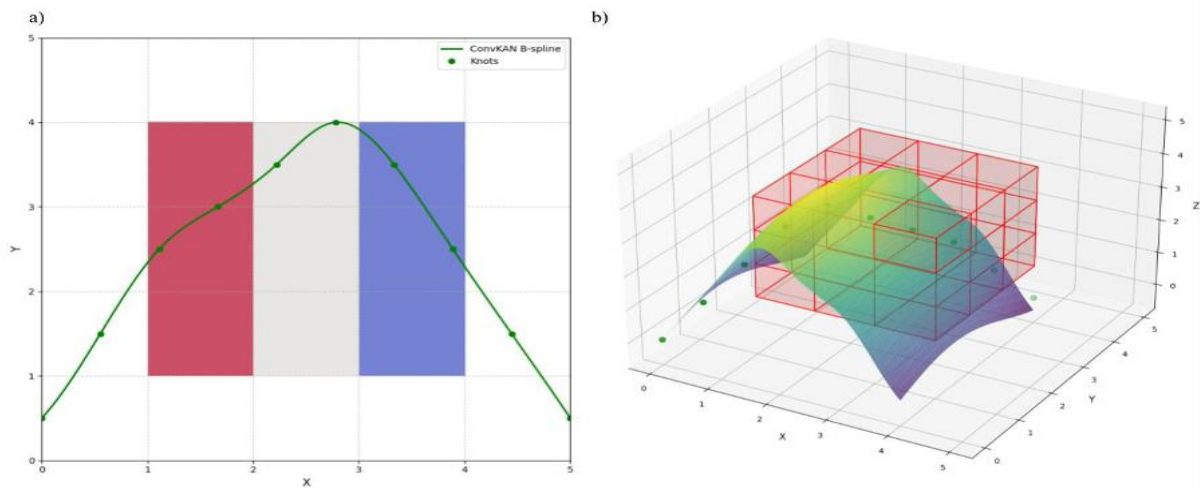

Figure 6: *Comparison of ConvKAN splines and CNN convolutional filters in 2D and 3D dimensions. (a) 2D representation: The gridded square represents a single MRI slice. The coloured 3x3 region within the grid illustrates an example CNN filter, while the green curve demonstrates a B-spline used in ConvKAN, with green dots indicating knots. (b) 3D representation: The cube represents a volumetric MRI scan. The red 3x3x3 region*

*within the cube shows a CNN filter, while the curved surface represents a 3D B-spline used in ConvKAN, with green dots marking the knots.*

We also implemented Graph Convolutional Network (GCN) models, transforming MRI data into graph representations. For the 2D GCN model, the Simple Linear Iterative Clustering (SLIC) algorithm was applied to generate 1000 superpixels per axial slice.[23] Node features included the mean intensity of each superpixel, relative area, and the centroid coordinates. In the 3D GCN model, we extended SLIC to create 1000 supervoxels per volumetric scan, using mean intensity, relative volume, and centroid coordinates as node features. Edges were constructed using a k-nearest neighbors (k=6) approach based on Euclidean distances between centroids.

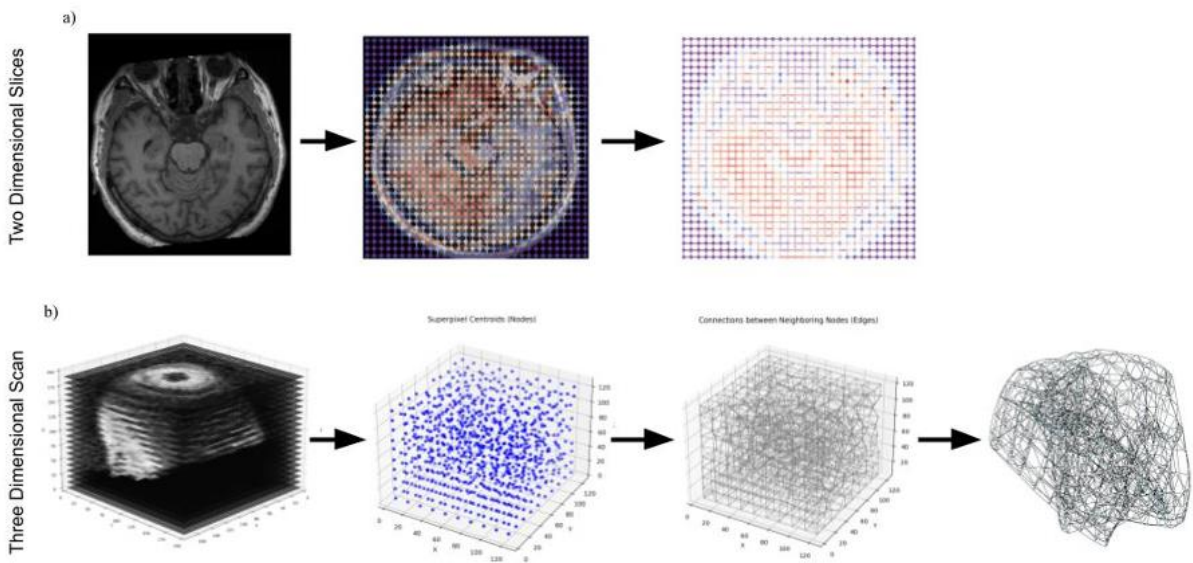

Figure 7: Graph creation processes for GCN analysis. a) 2D graph creations, using superpixels, demonstrating alignment with anatomical features. (b) 3D graph representation of volumetric MRI, with nodes as supervoxels and edges as spatial relationships. Each node (sphere) represents a supervoxel, with node size indicating relative volume and colour representing mean intensity. Edges (lines) connect neighbouring supervoxels in 3D space.

The GCN architecture for both 2D and 3D analysis comprised three graph convolutional layers followed by global mean pooling and two fully connected layers. For 2D GCN, the input was the superpixel graph derived from SLIC-processed slices. For 3D GCN, supervoxel graphs were used as inputs.

A baseline CNN architecture was also implemented, consisting of four convolutional layers with batch normalization and max pooling after each layer. Global average pooling was applied before two fully connected layers. Detailed descriptions of model architectures and graph creation methodology are provided in the Supplementary Material 2 and 3.

# Validation Strategies

## Isolated Dataset Analysis

To comprehensively evaluate the performance of our models, we employed two distinct validation strategies. First, we conducted an isolated dataset analysis using different cross-validation methods for 2D and 3D models. For 2D models, we implemented five-fold stratified group cross-validation, ensuring that all slices from a single subject were either in the training or validation set to prevent data leakage and maintain sample independence. For 3D models, given the smaller sample size of full-brain volumes, we used Leave-One-Out Cross-Validation (LOOCV) to maximize the use of limited data while maintaining unbiased evaluation.[24]

## Hold-Out Analysis

Second, we performed a hold-out analysis, also known as a "2 vs 1" analysis, to assess the generalizability of our models across different cohorts and scanning protocols. In this approach, we trained models on two datasets and tested them on the third, rotating through each dataset as the test set. This method was applied to both 2D and 3D models, allowing us to evaluate cross-dataset performance.

By utilizing these complementary validation strategies, we were able to assess both within-dataset performance and cross-dataset generalizability, providing a robust and comprehensive evaluation of our models' capabilities in various scenarios.

# Training and Evaluation

All models were trained using the Adam optimizer with an initial learning rate of 1e-4 and weight decay of 1e-5. Early stopping with a patience of 15 epochs was employed to avoid overfitting. To address class imbalance, we used weighted cross-entropy loss, with weights inversely proportional to the class frequencies in the training set.

Model performance was evaluated using accuracy, F1 score, recall, and Area Under the Receiver Operating Characteristic Curve (AUROC). To assess the uncertainty of performance estimates, 95% confidence intervals were computed using bootstrap resampling. To compare the performance of different models and the 2D vs. 3D implementations, we used the Wilcoxon signed-rank test and applied the Benjamini-Hochberg correction to control the false discovery rate (FDR) for multiple comparisons. P-values < 0.05 after FDR correction were considered statistically significant.

## Statistical Analysis

To comprehensively evaluate the performance and consistency of the 2D and 3D implementations of CNN, ConvKAN, and GCN models across multiple datasets, we employed a rigorous statistical analysis framework. This framework was designed to provide robust insights into model performance, generalizability, and statistical significance of observed differences.

For comparing the performance of different models and between 2D and 3D implementations, we primarily used the Wilcoxon signed-rank test. This non-parametric test was chosen over parametric alternatives (e.g., paired t-test) due to its robustness against non-normal distributions and outliers, which are common in machine learning performance metrics.

Prior to applying the Wilcoxon test, we used the Shapiro-Wilk test to assess the normality of the performance distributions. This step was included to ensure the appropriateness of our statistical approach and to identify any cases where parametric tests might be more suitable.

To control the false discovery rate in multiple comparisons, we applied the Benjamini-Hochberg procedure. This method was chosen over more conservative approaches (e.g., Bonferroni correction) to balance the control of Type I errors with maintaining statistical power.

This post-hoc statistical analysis was designed to provide a nuanced understanding of model performance beyond simple point estimates. The combination of non-parametric tests, effect size calculations, and bootstrap confidence intervals offers robustness against potential violations of parametric assumptions.

We calculated Cohen's d as a measure of effect size for all comparisons. This standardized measure allows for the assessment of the magnitude of differences between models or implementations, complementing the p-values from statistical tests.

Bootstrap resampling was used to compute 95% confidence intervals for all performance metrics and effect sizes. This non-parametric approach provides robust estimates of uncertainty, particularly valuable given the potentially non-normal distributions of our performance metrics.

To evaluate the consistency of model performance across datasets, we calculated the coefficient of variation (CV) for each model's performance metrics. The CV, being a standardized measure of dispersion, allows for fair comparisons of consistency across different metrics and models.

A power analysis was performed to justify the sample sizes, based on effect sizes observed in preliminary experiments. For an effect size of Cohen's d = 0.8 (large effect), an α of 0.05, and desired power of 0.8, the required sample size for each dataset was calculated. This analysis ensures that our study has sufficient statistical power to detect meaningful differences between models and datasets.

## Computational Resources and Data Availability

All models were trained using NVIDIA A100 GPUs via a publicly accessible cloud computing platform. Average training and inference times for each model were recorded and reported. The datasets used in this study are publicly available through their respective repositories: PPMI (https://www.ppmi-info.org), NEUROCON, and Tao Wu (https://fcon_1000.projects.nitrc.org/indi/retro/parkinsons.html). All code for model implementation and analysis will be made available on GitHub.

## Declaration of Interests

This research received no specific grant from any funding agency in the public, commercial, or not-for-profit sectors. SBP receives funding from the NIHR. VG has received grants from Siemens Healthineers and Lunit; honoraria from the European School of Radiology; and is Chair of the Royal College of Radiologists academic committee. JJF was supported by the NIHR Oxford Biomedical Research Centre. CAA receives funding from NIHR, UCB-Oxford collaboration, and Merck USA. These relationships are unrelated to the submitted work. The authors had full access to all study data and final responsibility for publication.

## Data Sharing

This study utilised three open-source datasets for model development and validation: the Parkinson's Progression Markers Initiative (PPMI) MRI dataset, the NEUROCON dataset, and the Tao Wu dataset. All datasets are publicly available and can be accessed through their respective online repositories: https://www.ppmi-info.org and https://fcon_1000.projects.nitrc.org/indi/retro/parkinsons.html. The code used for model implementation and analysis will be made available on GitHub.

# References

1. Ben-Shlomo Y, Darweesh S, Llibre-Guerra J, Marras C, San Luciano M, Tanner C. The epidemiology of Parkinson's disease. *Lancet*. 2024;**403**(10423):283-92.

2. Morris HR, Spillantini MG, Sue CM, Williams-Gray CH. The pathogenesis of Parkinson's disease. *Lancet*. 2024;**403**(10423):293-304.

3. Tolosa E, Garrido A, Scholz SW, Poewe W. Challenges in the diagnosis of Parkinson's disease. Lancet Neurol. 2021 May;20(5):385-397. doi: 10.1016/S1474-4422(21)00030-2. PMID: 33894193; PMCID: PMC8185633.

4. Beach TG, Adler CH. Importance of low diagnostic Accuracy for early Parkinson's disease. Mov Disord. 2018 Oct;33(10):1551-1554. doi: 10.1002/mds.27485. Epub 2018 Oct 4. PMID: 30288780; PMCID: PMC6544441.

5. Marsili L, Rizzo G, Colosimo C. Diagnostic Criteria for Parkinson's Disease: From James Parkinson to the Concept of Prodromal Disease. *Front Neurol*. 2018;**9**:156.

6.  Liu X, Faes L, Kale AU, et al. A comparison of deep learning performance against health-care professionals in detecting diseases from medical imaging: a systematic review and meta-analysis. *Lancet Digit Health.* 2019;**1**(6):e271-97.

7. Litjens G, Kooi T, Bejnordi BE, et al. A survey on deep learning in medical image analysis. Med Image Anal. 2017;42:60-88.

8. Mall PK, Singh PK, Srivastav S, Narayan V, Paprzycki M, Jaworska T, Ganzha M. A comprehensive review of deep neural networks for medical image processing: Recent developments and future opportunities. Healthc Anal. 2023;4:100216.

9. Esteva A, Robicquet A, Ramsundar B, et al. A guide to deep learning in healthcare. Nat Med. 2019;25(1):24-29.

10. Ahmedt-Aristizabal D, Armin MA, Denman S, Fookes C, Petersson L. Graph-Based Deep Learning for Medical Diagnosis and Analysis: Past, Present and Future. Sensors (Basel). 2021 Jul 12;21(14):4758. doi: 10.3390/s21144758. PMID: 34300498; PMCID: PMC8309939.

11. Multi-task graph structure learning based on node clustering for early Parkinson's disease diagnosis. *Comput Biol Med*. 2023;**152**:106308.

12. [preprint] Liu Z, Wang Y, Vaidya S, et al. KAN: Kolmogorov-Arnold Networks. arXiv [cs.LG]. 2024. Available from: https://arxiv.org/abs/2404.19756

13. .[preprint] Bodner AD, Tepsich AS, Spolski JN, Pourteau S. Convolutional Kolmogorov-Arnold Networks. arXiv [cs.CV]. 2024. Available from: https://arxiv.org/abs/2406.13155

14. [preprint] Bodner AD, Tepsich AS, Spolski JN, Pourteau S. Convolutional Kolmogorov-Arnold Networks. arXiv [cs.CV]. 2024. Available from: https://arxiv.org/abs/2406.13155

15. Ilesanmi AE, Ilesanmi TO, Ajayi BO. Reviewing 3D convolutional neural network approaches for medical image segmentation. Heliyon. 2024;10(6)

16. Starke S, Leger S, Zwanenburg A, et al. 2D and 3D convolutional neural networks for outcome modelling of locally advanced head and neck squamous cell carcinoma. Sci Rep. 2020;10:15625.

17. Tiwari S, Jain G, Shetty DK, Sudhi M, Balakrishnan JM, Bhatta SR. A comprehensive review on the application of 3D convolutional neural networks in medical imaging. Eng Proc. 2023;59(1):3.

18. Marek K, Chowdhury S, Siderowf A, et al. The Parkinson's progression markers initiative (PPMI) - establishing a PD biomarker cohort. *Ann Clin Transl Neurol*. 2018;**5**(12):1460-1477.

19. National Institute for Research and Development in Informatics, Romania; Department of Neurobiology, Beijing Institute of Geriatrics, Xuanwu Hospital, Capital Medical University; Parkinson Disease Centre of Beijing Institute for Brain Disorders, China. Parkinson's Disease Datasets. Available from: https://fcon_1000.projects.nitrc.org/indi/retro/parkinsons.html

20. 11. Huang L, Ye X, Yang M, Pan L, Zheng S. MNC-Net: Multi-task graph structure learning based on node clustering for early Parkinson's disease diagnosis. Comput Biol Med. 2023;152:106308.

21. Stern S, Lau S, Manole A, et al. Reduced synaptic activity and dysregulated extracellular matrix pathways in midbrain neurons from Parkinson's disease patients. *npj Parkinson's Disease*. 2022;**8:**103.

22. Kano H, Nakata H, Martin CF. Optimal curve fitting and smoothing using normalized uniform B-splines: a tool for studying complex systems. *Appl Math Comput.* 2005;**169**(1):96-128.

23. Achanta R, Shaji A, Smith K, Lucchi A, Fua P, Süsstrunk S. SLIC Superpixels Compared to State-of-the-art Superpixel Methods. *IEEE Trans Pattern Anal Mach Intell*. 2012;**34**(11):2274-82.

24. Vehtari A, Gelman A, Gabry J. Practical Bayesian model evaluation using leave-one-out cross-validation and WAIC. Stat Comput. 2017;27:1413-1432.

# Tables

| Parameter | PPMI | NEUROCON | Tao Wu |
|---|---|---|---|
| Total Participants | 59 | 43 | 40 |
| Age (SD) | 63.5 (11.1) | 68.3 (11.0) | 65.0 (5.0) |
| Number of PD | 28 | 27 | 20 |
| Number of Controls | 31 | 16 | 20 |
| Age PD (SD) | 61.6 (10.4) | 68.7 (11.0) | 65.2 (4.4) |
| Age Controls (SD) | 64.2 (8.7) | 67.6 (11.9) | 64.8 (5.6) |
| Disease Duration (years, SD) | 1.7 (0.8) | 4.8 (6.2) | 5.4 (3.9) |
| Number of Males | 26 | 21 | 23 |
| Number of Females | 33 | 22 | 17 |
| MRI Sequence (T1) | MPRAGE (T1) | MPRAGE (T1) | MPRAGE (T1) |
| TR (ms) | * | 1940 | 1100 |
| TE (ms) | * | 3.08 | 3.39 |
| Voxel Size (mm) | * | 0.97 x 0.97 x 1.0 | 1.0 x 1.0 x 1.0 |

Table 1: Participant demographics and MRI parameters for each dataset. * PPMI scans are variable regarding sequence parameters.

# Supplementary Material

# 1. Supplementary Results

This section provides a detailed presentation of all results from our study on MRI-based Parkinson's Disease classification using various deep learning architectures. While some of these results are discussed in the main manuscript, we present here a more comprehensive view, including full statistical analyses, detailed performance metrics, and additional evaluations.

## 1. Isolated Dataset Analysis

### 1.1 Full Performance Metrics

Table S1 presents the complete performance metrics for all models across the three datasets (PPMI, Tao Wu, and NEUROCON) in the isolated dataset analysis. Metrics include accuracy, F1 score, recall, and Area Under the Receiver Operating Characteristic Curve (AUC), with 95% confidence intervals.

Table S1: Full Performance Metrics for Isolated Dataset Analysis

| Dataset | Model | Accuracy (95% CI) | F1 Score (95% CI) | Recall (95% CI) | AUC (95% CI) |
|---|---|---|---|---|---|
| **PPMI** | 2D CNN | 0.90 (0.89-0.91) | 0.90 (0.89-0.91) | 0.90 (0.89-0.91) | 0.97 (0.96-0.98) |
| | 2D ConvKAN | 0.93 (0.92-0.94) | 0.93 (0.92-0.94) | 0.93 (0.92-0.94) | 0.99 (0.98-0.99) |
| | 2D GCN | 0.53 (0.52-0.54) | 0.33 (0.32-0.34) | 0.53 (0.52-0.54) | 0.54 (0.53-0.55) |
| | 3D CNN | 0.85 (0.84-0.86) | 0.84 (0.83-0.85) | 0.85 (0.84-0.86) | 0.93 (0.92-0.94) |
| | 3D ConvKAN | 0.83 (0.82-0.84) | 0.82 (0.81-0.83) | 0.83 (0.82-0.84) | 0.85 (0.84-0.86) |
|  | 3D GCN | 0.83 (0.82-0.84) | 0.83 (0.82-0.84) | 0.83 (0.82-0.84) | 0.93 (0.92-0.94) |
| **Tao Wu** | 2D CNN | 0.84 (0.83-0.85) | 0.84 (0.83-0.85) | 0.84 (0.83-0.85) | 0.90 (0.89-0.91) |
|  | 2D ConvKAN | 0.84 (0.83-0.85) | 0.84 (0.83-0.85) | 0.84 (0.83-0.85) | 0.93 (0.92-0.94) |
|  | 2D GCN | 0.50 (0.49-0.51) | 0.33 (0.32-0.34) | 0.50 (0.49-0.51) | 0.46 (0.45-0.47) |
|  | 3D CNN | 0.93 (0.92-0.94) | 0.91 (0.90-0.92) | 0.93 (0.92-0.94) | 0.94 (0.93-0.95) |
|  | 3D ConvKAN | 0.88 (0.87-0.89) | 0.87 (0.86-0.88) | 0.88 (0.87-0.89) | 0.91 (0.90-0.92) |
|  | 3D GCN | 0.60 (0.59-0.61) | 0.48 (0.47-0.49) | 0.60 (0.59-0.61) | 0.50 (0.49-0.51) |
| **NEUROCON** | 2D CNN | 0.75 (0.74-0.76) | 0.75 (0.74-0.76) | 0.75 (0.74-0.76) | 0.80 (0.79-0.81) |
|  | 2D ConvKAN | 0.75 (0.74-0.76) | 0.74 (0.73-0.75) | 0.75 (0.74-0.76) | 0.84 (0.83-0.85) |
|  | 2D GCN | 0.63 (0.62-0.64) | 0.48 (0.47-0.49) | 0.63 (0.62-0.64) | 0.55 (0.54-0.56) |
|  | 3D CNN | 0.89 (0.88-0.90) | 0.89 (0.88-0.90) | 0.89 (0.88-0.90) | 0.90 (0.89-0.91) |
|  | 3D ConvKAN | 0.86 (0.85-0.87) | 0.84 (0.83-0.85) | 0.86 (0.85-0.87) | 0.91 (0.90-0.92) |
|  | 3D GCN | 0.83 (0.82-0.84) | 0.83 (0.82-0.84) | 0.83 (0.82-0.84) | 0.89 (0.88-0.90) |

1.2 Statistical Comparisons

We performed comprehensive statistical comparisons between models and dimensionalities. The following subsections present these comparisons, with p-values corrected for multiple comparisons using the Benjamini-Hochberg false discovery rate (FDR) method.

## 1.2.1 2D vs 3D Model Comparisons

Table S2 presents the statistical comparisons between 2D and 3D versions of each model type across all datasets and metrics.

Table S2: 2D vs 3D Model Comparisons for Isolated Dataset Analysis

| Dataset | Comparison | Metric | p-value | Corrected p-value | Effect Size |
| --- | --- | --- | --- | --- | --- |
| PPMI | 2D CNN vs 3D CNN | Accuracy | 0.5947 | 0.7382 | 0.39 |
| | | F1 Score | 0.5671 | 0.7382 | 0.42 |
| | | Recall | 0.5947 | 0.7382 | 0.39 |
| | | AUC | 0.8125 | 0.9181 | 0.63 |
| | 2D ConvKAN vs 3D ConvKAN | Accuracy | 0.0452 | 0.1099 | 1.95 |
| | | F1 Score | 0.0648 | 0.1099 | 1.70 |
| | | Recall | 0.0452 | 0.1099 | 1.95 |
| | | AUC | 0.8125 | 0.9181 | 0.63 |
| | 2D GCN vs 3D GCN | Accuracy | 0.0625 | 0.1099 | -3.53 |
| | | F1 Score | 0.0625 | 0.1099 | -5.17 |
| | | Recall | 0.0625 | 0.1099 | -3.53 |
| | | AUC | 0.0625 | 0.1099 | -2.90 |
| Tao Wu | 2D CNN vs 3D CNN | Accuracy | 0.1875 | 0.3009 | -1.11 |
| | | F1 Score | 0.1875 | 0.3009 | -0.92 |
| | | Recall | 0.1875 | 0.3009 | -1.11 |
| | | AUC | 0.1875 | 0.3009 | -0.76 |
| | 2D ConvKAN vs 3D ConvKAN | Accuracy | 0.6250 | 0.6250 | -0.15 |
| | | F1 Score | 0.6250 | 0.6250 | -0.15 |
| | | Recall | 0.6250 | 0.6250 | -0.15 |
| | | AUC | 0.6250 | 0.6250 | 0.18 |
| | 2D GCN vs 3D GCN | Accuracy | 0.3739 | 0.4206 | -0.63 |
| | | F1 Score | 0.1100 | 0.2201 | -1.29 |
| | | Recall | 0.3739 | 0.4206 | -0.63 |
| | | AUC | 0.3125 | 0.3808 | -0.99 |
| NEUROCON | 2D CNN vs 3D CNN | Accuracy | 0.3125 | 0.4027 | -0.69 |
| | | F1 Score | 0.3125 | 0.4027 | -0.70 |
| | | Recall | 0.3125 | 0.4027 | -0.69 |
| | | AUC | 0.1875 | 0.3257 | -0.49 |
| | 2D ConvKAN vs 3D ConvKAN | Accuracy | 0.6250 | 0.6875 | -0.31 |
| | | F1 Score | 0.8125 | 0.8125 | -0.25 |

| | | Recall | 0.6250 | 0.6875 | -0.31 |
| | | AUC | 0.1250 | 0.3257 | -0.40 |
| | 2D GCN vs 3D GCN | Accuracy | N/A | N/A | N/A |
| | | F1 Score | N/A | N/A | N/A |
| | | Recall | N/A | N/A | N/A |
| | | AUC | 0.5546 | 0.6536 | -0.42 |

Note: N/A values indicate that no statistical test could be performed due to identical values between 2D and 3D models.

Table S3: 2D Model Comparisons for Isolated Dataset Analysis

| Dataset | Comparison | Metric | p-value | Corrected p-value | Effect Size |
|---|---|---|---|---|---|
| PPMI | CNN vs ConvKAN | Accuracy | 0.0075 | 0.0539 | -1.26 |
| | | F1 Score | 0.0075 | 0.0539 | -1.26 |
| | | Recall | 0.0075 | 0.0539 | -1.26 |
| | | AUC | 0.0092 | 0.0550 | -1.13 |
| | CNN vs GCN | Accuracy | 0.0625 | 0.1099 | 11.88 |
|  |  | F1 Score | 0.0625 | 0.1099 | 15.32 |
|  |  | Recall | 0.0625 | 0.1099 | 11.88 |
|  |  | AUC | 0.0030 | 0.0539 | 4.17 |
|  | ConvKAN vs GCN | Accuracy | 0.0625 | 0.1099 | 11.80 |
|  |  | F1 Score | 0.0625 | 0.1099 | 15.08 |
|  |  | Recall | 0.0625 | 0.1099 | 11.80 |
|  |  | AUC | 0.0027 | 0.0539 | 4.32 |
| Tao Wu | CNN vs ConvKAN | Accuracy | 0.2090 | 0.3009 | -0.68 |
|  |  | F1 Score | 0.2032 | 0.3009 | -0.68 |
|  |  | Recall | 0.2090 | 0.3009 | -0.68 |
|  |  | AUC | 0.0342 | 0.1438 | -1.06 |
|  | CNN vs GCN | Accuracy | 0.0042 | 0.0253 | 3.71 |

| | | F1 Score | 0.0007 | 0.0129 | 5.94 |
| | | Recall | 0.0042 | 0.0253 | 3.71 |
| | | AUC | 0.0625 | 0.1438 | 3.73 |
| | ConvKAN vs GCN | Accuracy | 0.0022 | 0.0200 | 4.41 |
| | | F1 Score | 0.0005 | 0.0129 | 6.49 |
| | | Recall | 0.0022 | 0.0200 | 4.41 |
| | | AUC | 0.0625 | 0.1438 | 5.69 |
| NEUROCON | CNN vs ConvKAN | Accuracy | 0.7313 | 0.7542 | 0.06 |
| | | F1 Score | 0.5110 | 0.6245 | 0.08 |
| | | Recall | 0.7313 | 0.7542 | 0.06 |
| | | AUC | 0.3125 | 0.4027 | -0.27 |

| | CNN vs GCN | Accuracy | 0.1875 | 0.3257 | 1.34 |
| | | F1 Score | 0.0625 | 0.3257 | 2.58 |
| | | Recall | 0.1875 | 0.3257 | 1.34 |
| | | AUC | 0.0247 | 0.3257 | 2.62 |
| | ConvKAN vs GCN | Accuracy | 0.3125 | 0.4027 | 1.50 |
| | | F1 Score | 0.0625 | 0.3257 | 2.85 |
| | | Recall | 0.3125 | 0.4027 | 1.50 |
| | | AUC | 0.0625 | 0.3257 | 3.91 |

Table S4: 3D Model Comparisons for Isolated Dataset Analysis

| **Dataset** | **Comparison** | **Metric** | **p-value** | **Corrected p-value** | **Effect Size** |
|---|---|---|---|---|---|
| PPMI | CNN vs ConvKAN | Accuracy | 0.0672 | 0.1099 | -0.67 |
| | | F1 Score | 0.0339 | 0.1099 | -0.76 |
| | | Recall | 0.0672 | 0.1099 | -0.67 |
| | | AUC | 0.5930 | 0.7382 | -0.10 |
| | CNN vs GCN | Accuracy | 0.8416 | 0.9181 | 0.12 |
| | | F1 Score | 0.8779 | 0.9295 | 0.10 |
| | | Recall | 0.8416 | 0.9181 | 0.12 |
| | | AUC | 1.0000 | 1.0000 | -0.14 |
| | ConvKAN vs GCN | Accuracy | 0.3739 | 0.5384 | -0.68 |
| | | F1 Score | 0.3166 | 0.4956 | -0.79 |
| | | Recall | 0.3739 | 0.5384 | -0.68 |
| | | AUC | 1.0000 | 1.0000 | -0.21 |
| Tao Wu | CNN vs ConvKAN | Accuracy | 0.3173 | 0.3808 | -0.22 |
| | | F1 Score | 0.3173 | 0.3808 | -0.16 |
| | | Recall | 0.3173 | 0.3808 | -0.22 |
| | | AUC | 0.3173 | 0.3808 | -0.20 |
| | CNN vs GCN | Accuracy | 0.0625 | 0.1438 | 1.64 |
| | | F1 Score | 0.0625 | 0.1438 | 1.57 |
| | | Recall | 0.0625 | 0.1438 | 1.64 |
| | | AUC | 0.0625 | 0.1438 | 1.51 |
| | ConvKAN vs GCN | Accuracy | 0.0679 | 0.1438 | 1.09 |
| | | F1 Score | 0.0679 | 0.1438 | 1.17 |
| | | Recall | 0.0679 | 0.1438 | 1.09 |

| | | AUC | 0.0625 | 0.1438 | 1.03 |
| NEUROCON | CNN vs ConvKAN | Accuracy | 0.1797 | 0.3257 | -0.27 |
| | | F1 Score | 0.1797 | 0.3257 | -0.31 |
| | | Recall | 0.1797 | 0.3257 | -0.27 |
| | | AUC | 0.3173 | 0.4027 | 0.05 |
| | CNN vs GCN | Accuracy | 0.1250 | 0.3257 | 1.44 |
| | | F1 Score | 0.1250 | 0.3257 | 2.18 |
| | | Recall | 0.1250 | 0.3257 | 1.44 |
| | | AUC | 0.1250 | 0.3257 | 2.15 |
| | ConvKAN vs GCN | Accuracy | 0.1875 | 0.3257 | 0.90 |
| | | F1 Score | 0.1875 | 0.3257 | 1.41 |
| | | Recall | 0.1875 | 0.3257 | 0.90 |
| | | AUC | 0.0679 | 0.3257 | 2.40 |

## 1.3 Model Consistency Analysis

To assess the consistency of model performance across datasets, we calculated the coefficient of variation (CV) for each model's performance metrics. Lower CV values indicate higher consistency. Table S5 presents these results.

Table S5: Model Consistency Analysis for Isolated Dataset Analysis

| **Dimension** | **Model** | **Accuracy CV (%)** | **F1 Score CV (%)** | **Recall CV (%)** | **AUC CV (%)** |
| --- | --- | --- | --- | --- | --- |
| 2D | CNN | 7.94 | 8.05 | 7.94 | 8.23 |
| | ConvKAN | 8.90 | 9.17 | 8.90 | 6.40 |
| | GCN | 9.79 | 16.44 | 9.79 | 7.74 |
| 3D | CNN | 3.45 | 3.24 | 3.45 | 1.74 |
| | ConvKAN | 6.36 | 8.69 | 6.36 | 0.88 |
| | GCN | 15.15 | 23.73 | 15.15 | 24.74 |

## 1.4 Effect Size Analysis

We calculated Cohen's d effect sizes for the performance differences between 3D and 2D implementations of each model type. Table S6 presents these results with 95% confidence intervals.

Table S6: Effect Size Analysis for 3D vs 2D Model Comparisons in Isolated Dataset Analysis

| Dataset | Model | Accuracy Effect Size (95% CI) | F1 Score Effect Size (95% CI) | Recall Effect Size (95% CI) | AUC Effect Size (95% CI) |
|---|---|---|---|---|---|
| PPMI | CNN | 0.39 (-0.84 to 1.62) | 0.42 (-0.81 to 1.65) | 0.39 (-0.84 to 1.62) | 0.63 (-0.61 to 1.87) |
| | ConvKAN | 1.95 (0.59 to 3.31) | 1.70 (0.37 to 3.03) | 1.95 (0.59 to 3.31) | 0.63 (-0.61 to 1.87) |
| | GCN | -3.53 (-5.11 to -1.95) | -5.17 (-7.08 to -3.26) | -3.53 (-5.11 to -1.95) | -2.90 (-4.38 to -1.42) |
| Tao Wu | CNN | -1.11 (-2.39 to 0.17) | -0.92 (-2.18 to 0.34) | -1.11 (-2.39 to 0.17) | -0.76 (-2.00 to 0.48) |

| | ConvKAN | -0.15 (-1.35 to 1.05) | -0.15 (-1.35 to 1.05) | -0.15 (-1.35 to 1.05) | 0.18 (-1.02 to 1.38) |
| | GCN | -0.63 (-1.86 to 0.60) | -1.29 (-2.58 to 0.00) | -0.63 (-1.86 to 0.60) | -0.99 (-2.25 to 0.27) |
| NEUROCON | CNN | -0.69 (-1.93 to 0.55) | -0.70 (-1.94 to 0.54) | -0.69 (-1.93 to 0.55) | -0.49 (-1.71 to 0.73) |
| | ConvKAN | -0.31 (-1.52 to 0.90) | -0.25 (-1.46 to 0.96) | -0.31 (-1.52 to 0.90) | -0.40 (-1.62 to 0.82) |
| | GCN | N/A | N/A | N/A | -0.42 (-1.64 to 0.80) |

Note: N/A values indicate that effect sizes could not be calculated due to identical values between 2D and 3D models.

## 2. Hold-Out Analysis

### 2.1 Full Performance Metrics

Table S7 presents the complete performance metrics for all models across the three datasets in the hold-out analysis.

Table S7: Full Performance Metrics for Hold-Out Analysis

| Dataset | Model | Accuracy (95% CI) | F1 Score (95% CI) | Recall (95% CI) | AUC (95% CI) |
|---|---|---|---|---|---|
| PPMI | 2D CNN | 0.70 (0.69-0.71) | 0.70 (0.69-0.71) | 0.70 (0.69-0.71) | 0.79 (0.78-0.80) |

|  | 2D ConvKAN | 0.59 (0.58-0.60) | 0.54 (0.53-0.55) | 0.59 (0.58-0.60) | 0.54 (0.53-0.55) |
|---|---|---|---|---|---|
|  | 2D GCN | 0.47 (0.46-0.48) | 0.31 (0.30-0.32) | 0.47 (0.46-0.48) | 0.58 (0.57-0.59) |
|  | 3D CNN | 0.76 (0.75-0.77) | 0.75 (0.74-0.76) | 0.76 (0.75-0.77) | 0.80 (0.79-0.81) |
|  | 3D ConvKAN | 0.83 (0.82-0.84) | 0.82 (0.81-0.83) | 0.83 (0.82-0.84) | 0.85 (0.84-0.86) |
|  | 3D GCN | 0.69 (0.68-0.70) | 0.69 (0.68-0.70) | 0.69 (0.68-0.70) | 0.81 (0.80-0.82) |
| Tao Wu | 2D CNN | 0.57 (0.56-0.58) | 0.49 (0.48-0.50) | 0.57 (0.56-0.58) | 0.74 (0.73-0.75) |
|  | 2D ConvKAN | 0.57 (0.56-0.58) | 0.52 (0.51-0.53) | 0.57 (0.56-0.58) | 0.64 (0.63-0.65) |
|  | 2D GCN | 0.50 (0.49-0.51) | 0.33 (0.32-0.34) | 0.50 (0.49-0.51) | 0.55 (0.54-0.56) |
| | 3D CNN | 0.50 (0.49-0.51) | 0.33 (0.32-0.34) | 0.50 (0.49-0.51) | 0.72 (0.71-0.73) |
| | 3D ConvKAN | 0.65 (0.64-0.66) | 0.62 (0.61-0.63) | 0.65 (0.64-0.66) | 0.66 (0.65-0.67) |
| | 3D GCN | 0.63 (0.62-0.64) | 0.56 (0.55-0.57) | 0.63 (0.62-0.64) | 0.66 (0.65-0.67) |
| NEUROCON | 2D CNN | 0.57 (0.56-0.58) | 0.54 (0.53-0.55) | 0.57 (0.56-0.58) | 0.87 (0.86-0.88) |
| | 2D ConvKAN | 0.42 (0.41-0.43) | 0.30 (0.29-0.31) | 0.42 (0.41-0.43) | 0.82 (0.81-0.83) |
| | 2D GCN | 0.63 (0.62-0.64) | 0.48 (0.47-0.49) | 0.63 (0.62-0.64) | 0.46 (0.45-0.47) |
| | 3D CNN | 0.63 (0.62-0.64) | 0.48 (0.47-0.49) | 0.63 (0.62-0.64) | 0.48 (0.47-0.49) |
| | 3D ConvKAN | 0.63 (0.62-0.64) | 0.48 (0.47-0.49) | 0.63 (0.62-0.64) | 0.48 (0.47-0.49) |

| | 3D GCN | 0.63 (0.62-0.64) | 0.48 (0.47-0.49) | 0.63 (0.62-0.64) | 0.56 (0.55-0.57) |

## 2.2 Statistical Comparisons

Due to the limited number of data points in the hold-out analysis (one per model per dataset), traditional statistical tests were not applicable. Instead, we present descriptive comparisons of the performance differences.

### 2.2.1 2D vs 3D Model Comparisons

Table S8 presents the performance differences between 2D and 3D versions of each model type across all datasets and metrics for the hold-out analysis.

Table S8: 2D vs 3D Model Performance Differences in Hold-Out Analysis

| Dataset | Comparison | Accuracy Diff | F1 Score Diff | Recall Diff | AUC Diff |
|---|---|---|---|---|---|
| PPMI | 3D CNN - 2D CNN | 0.06 | 0.05 | 0.06 | 0.01 |
| | 3D ConvKAN - 2D ConvKAN | 0.24 | 0.28 | 0.24 | 0.31 |
| | 3D GCN - 2D GCN | 0.22 | 0.38 | 0.22 | 0.23 |
| Tao Wu | 3D CNN - 2D CNN | -0.07 | -0.16 | -0.07 | -0.02 |

|  | 3D ConvKAN - 2D ConvKAN | 0.08 | 0.10 | 0.08 | 0.02 |
|  | 3D GCN - 2D GCN | 0.13 | 0.23 | 0.13 | 0.11 |
| NEUROCON | 3D CNN - 2D CNN | 0.06 | -0.06 | 0.06 | -0.39 |
|  | 3D ConvKAN - 2D ConvKAN | 0.21 | 0.18 | 0.21 | -0.34 |
|  | 3D GCN - 2D GCN | 0.00 | 0.00 | 0.00 | 0.10 |

## 2.2.2 Within-Dimension Model Comparisons

Tables S9 and S10 present the performance differences between different model architectures within the 2D and 3D dimensions, respectively, for the hold-out analysis.

Table S9: 2D Model Performance Differences in Hold-Out Analysis

| Dataset | Comparison | Accuracy Diff | F1 Score Diff | Recall Diff | AUC Diff |
|---|---|---|---|---|---|
| PPMI | CNN - ConvKAN | 0.11 | 0.16 | 0.11 | 0.25 |
|  | CNN - GCN | 0.23 | 0.39 | 0.23 | 0.21 |

|  | Comparison | Accuracy Diff | F1 Score Diff | Recall Diff | AUC Diff |
|---|---|---|---|---|---|
|  | ConvKAN - GCN | 0.12 | 0.23 | 0.12 | -0.04 |
| Tao Wu | CNN - ConvKAN | 0.00 | -0.03 | 0.00 | 0.10 |
|  | CNN - GCN | 0.07 | 0.16 | 0.07 | 0.19 |
|  | ConvKAN - GCN | 0.07 | 0.19 | 0.07 | 0.09 |
| NEUROCON | CNN - ConvKAN | 0.15 | 0.24 | 0.15 | 0.05 |
|  | CNN - GCN | -0.06 | 0.06 | -0.06 | 0.41 |
|  | ConvKAN - GCN | -0.21 | -0.18 | -0.21 | 0.36 |

Table S10: 3D Model Performance Differences in Hold-Out Analysis

| Dataset | Comparison | Accuracy Diff | F1 Score Diff | Recall Diff | AUC Diff |
|---|---|---|---|---|---|
| PPMI | CNN - ConvKAN | -0.07 | -0.07 | -0.07 | -0.05 |
|  | CNN - GCN | 0.07 | 0.06 | 0.07 | -0.01 |
|  | ConvKAN - GCN | 0.14 | 0.13 | 0.14 | 0.04 |

| Tao Wu | CNN - ConvKAN | -0.15 | -0.29 | -0.15 | 0.06 |
| | CNN - GCN | -0.13 | -0.23 | -0.13 | 0.06 |
| | ConvKAN - GCN | 0.02 | 0.06 | 0.02 | 0.00 |
| NEUROCON | CNN - ConvKAN | 0.00 | 0.00 | 0.00 | 0.00 |
| | CNN - GCN | 0.00 | 0.00 | 0.00 | -0.08 |
| | ConvKAN - GCN | 0.00 | 0.00 | 0.00 | -0.08 |

## 2.3 Model Consistency Analysis

Table S11 presents the coefficient of variation (CV) for each model's performance metrics across datasets in the hold-out analysis.

Table S11: Model Consistency Analysis for Hold-Out Analysis

| Dimension | Model | Accuracy CV (%) | F1 Score CV (%) | Recall CV (%) | AUC CV (%) |
|---|---|---|---|---|---|
| 2D | CNN | 10.22 | 15.45 | 10.22 | 6.79 |
| | ConvKAN | 14.59 | 24.48 | 14.59 | 17.32 |

|  | GCN | 12.56 | 20.99 | 12.56 | 10.06 |
|---|---|---|---|---|---|
| 3D | CNN | 17.02 | 32.64 | 17.02 | 20.40 |
|  | ConvKAN | 12.91 | 21.80 | 12.91 | 23.05 |
|  | GCN | 4.97 | 14.44 | 4.97 | 15.18 |

## 2.4 Effect Size Analysis

Due to the limited number of data points in the hold-out analysis, effect sizes could not be calculated with confidence intervals. Instead, we present point estimates of Cohen's d effect sizes for the performance differences between 3D and 2D implementations of each model type in Table S12.

Table S12: Effect Size Estimates for 3D vs 2D Model Comparisons in Hold-Out Analysis

| Dataset | Model | Accuracy Effect Size | F1 Score Effect Size | Recall Effect Size | AUC Effect Size |
|---|---|---|---|---|---|
| PPMI | CNN | 0.39 | 0.32 | 0.39 | 0.06 |
|  | ConvKAN | 1.95 | 1.70 | 1.95 | 1.93 |
|  | GCN | 1.45 | 2.04 | 1.45 | 1.40 |

| Tao Wu | CNN | -0.45 | -0.81 | -0.45 | -0.12 |
|---|---|---|---|---|---|
| | ConvKAN | 0.52 | 0.51 | 0.52 | 0.12 |
| | GCN | 0.84 | 1.18 | 0.84 | 0.67 |
| NEUROCON | CNN | 0.39 | -0.30 | 0.39 | -2.37 |
| | ConvKAN | 1.35 | 0.92 | 1.35 | -2.06 |
| | GCN | 0.00 | 0.00 | 0.00 | 0.61 |

These comprehensive results provide a detailed view of model performance across different datasets and analysis scenarios, highlighting the strengths and limitations of each approach in the context of MRI-based Parkinson's Disease classification.

# 2. Model Architectures and Rationale

We implemented three distinct neural network architectures - Convolutional Neural Networks (CNNs), Convolutional Kolmogorov-Arnold Networks (ConvKANs), and Graph Convolutional Networks (GCNs) - each in both 2D and 3D variations to process slice-based and volumetric MRI data, respectively.

Our 2D models process individual MRI slices of dimensions 224x224 pixels, extracted from the central region of the brain encompassing 100 axial slices. The 3D models operate on entire brain volumes resampled to 128x128x128 voxels. This dimensional choice balances spatial resolution with computational feasibility, allowing for feature extraction while managing memory constraints.

# Convolutional Neural Networks (CNNs)

## 2D CNN Architecture

The 2D CNN comprises four convolutional layers with 32, 64, 128, and 256 filters respectively, each using 3x3 kernels with stride 1 and padding 1. Each convolutional layer is followed by batch normalization (momentum 0.1, epsilon 1e-5) and max pooling (2x2). The network concludes with an adaptive average pooling layer to produce a 1x1 spatial output, followed by two fully connected layers: the first with 64 units and the second with 2 units for binary classification.

Mathematically, each convolutional layer performs the following operation:

$$y\_l = Pool(BN(ReLU(W\_l * x\_l + b\_l)))$$

where x_l is the input to layer l, W_l and b_l are the learnable weights and biases, BN denotes batch normalization, and Pool is the max pooling operation.

## 3D CNN Architecture

The 3D CNN extends this architecture to volumetric data, employing four 3D convolutional layers with 32, 64, 128, and 256 filters, each using 3x3x3 kernels. Similar to its 2D counterpart, each convolutional layer is succeeded by batch normalization and 3D max pooling (2x2x2). The network culminates in an adaptive 3D average pooling layer and two fully connected layers (64 units, then 2 units).

Both CNN variants utilize ReLU activation functions and incorporate dropout (p=0.5) before the final classification layer to mitigate overfitting.

# Convolutional Kolmogorov-Arnold Networks (ConvKANs)

ConvKANs represent an innovative fusion of Kolmogorov-Arnold Network principles with convolutional architectures. The core innovation lies in the integration of learnable B-spline functions as adaptive, non-linear activations within the convolutional framework, offering enhanced flexibility in modeling complex relationships in MRI data.

## 2D ConvKAN Implementation

Our 2D ConvKAN utilizes custom SplineConv2d layers. Each SplineConv2d layer consists of:

1. A standard 2D convolution operation.
2. A BSpline2D module with 10 knots and 6 control points.
3. A hybrid activation mechanism combining the B-spline output with a SiLU (Sigmoid Linear Unit) activation.

The forward pass of a SplineConv2d layer can be described as:

output = w1 * spline(conv(x)) + w2 * silu(conv(x))

where w1 and w2 are learnable parameters, spline() is the B-spline function, and conv() is the standard convolution operation.

The B-spline function is defined as:

$B(x) = \Sigma(i=0 \text{ to } n) c\_i * N\_{i,p}(x)$

where $c\_i$ are the control points, $N\_{i,p}$ are the basis functions of degree p, and n is the number of control points minus one.

The 2D ConvKAN architecture comprises:

- Three SplineConv2d layers with 32, 64, and 128 filters respectively (3x3 kernel, stride 1, padding 1)
- Each SplineConv2d layer is followed by batch normalization and max pooling (2x2)
- An adaptive average pooling layer
- Two fully connected layers (64 units, then 2 units for binary classification)

## 3D ConvKAN Implementation

The 3D ConvKAN extends this concept to volumetric data using KANConv3D layers. Each KANConv3D layer includes:

1. A standard 3D convolution operation.
2. A BSpline3D module, again with 10 knots and 6 control points, adapted for 3D inputs.
3. The same hybrid activation mechanism as in 2D, but applied to 3D data.

The 3D ConvKAN structure consists of:

- Four KANConv3D layers with 32, 64, 128, and 256 filters (3x3x3 kernel, stride 1, padding 1)
- Each KANConv3D layer is followed by 3D batch normalization and max pooling (2x2x2)
- A 3D adaptive average pooling layer
- Two fully connected layers (64 units, then 2 units for classification)

In both 2D and 3D implementations, the B-spline functions allow for dynamic adaptation of the activation landscape during training. This adaptability enables the capture of more nuanced, non-linear features in the MRI data compared to traditional CNNs with fixed activation functions.

A key feature of both implementations is the grid extension mechanism. When input values fall outside the initial B-spline range, the grid is dynamically extended by 25% in the required direction, with spline coefficients interpolated to maintain continuity. This ensures that the B-spline functions can effectively handle a wide range of input values encountered during training and inference.

The combination of convolutional operations, adaptive B-spline activations, and the hybrid mechanism with SiLU activation provides a flexible framework for capturing complex spatial relationships in both 2D slices and 3D volumes of MRI data, offering advantages in detecting subtle structural changes associated with Parkinson's Disease.

# Graph Convolutional Networks (GCNs)

Our GCN models offer a fundamentally different approach to MRI analysis, representing the image data as graphs. This approach allows for the explicit modeling of structural relationships within the brain, which may be particularly relevant for detecting the subtle anatomical changes associated with PD.

For 2D analysis, each MRI slice is transformed into a graph with 1000 nodes, where each node represents a superpixel. Node features include mean intensity and relative size of the superpixel. In the 3D case, we create a single graph per volume using 1000 supervoxels, with node features encompassing mean intensity, relative volume, and normalized centroid coordinates.

Edge connections in both 2D and 3D graphs are established using a k-nearest neighbors approach (k=6) based on the Euclidean distance between centroids. Edge weights are computed as the inverse of the distance between connected nodes.

The 2D GCN architecture comprises four GCNConv layers with 64 hidden channels each. The 3D GCN maintains the same structure with four GCNConv layers and 64 hidden channels. Both variants employ ReLU activation and dropout (p=0.3) after each GCNConv layer, followed by global mean pooling and a final linear layer for classification.

Mathematically, each GCNConv layer performs the following operation:

$$H^{(l+1)} = \sigma(D^{-1/2} \tilde{A} D^{-1/2} H^{(l)} W^{(l)})$$

where $H^{(l)}$ is the node feature matrix at layer $l$, $\tilde{A}$ is the adjacency matrix with self-loops, $D$ is the degree matrix, $W^{(l)}$ is the learnable weight matrix, and $\sigma$ is the ReLU activation function.

This graph-based approach allows our models to capture both local and global structural information in the MRI data, theoretically identifying patterns that might be overlooked by traditional convolutional approaches.

a)  b)

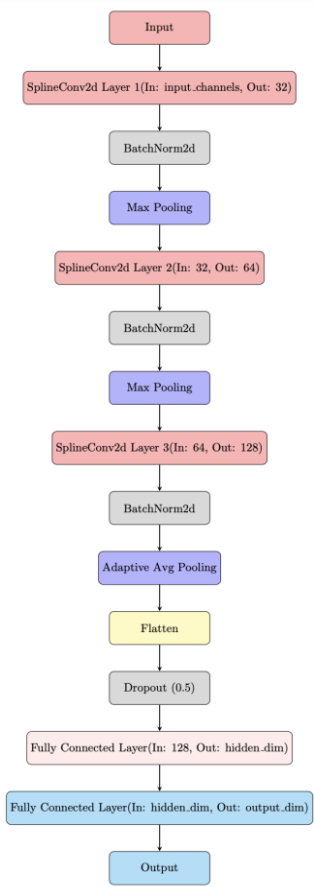 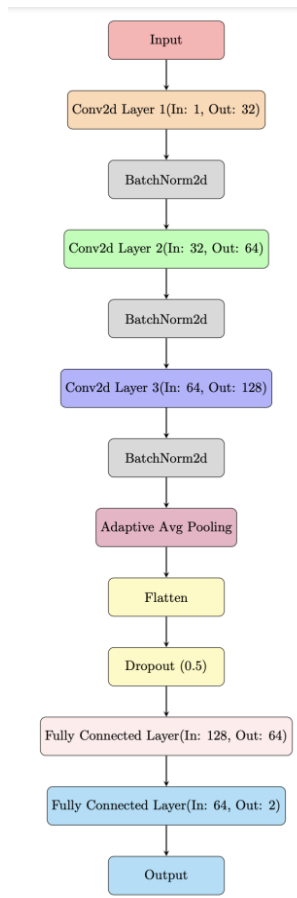

c) d)

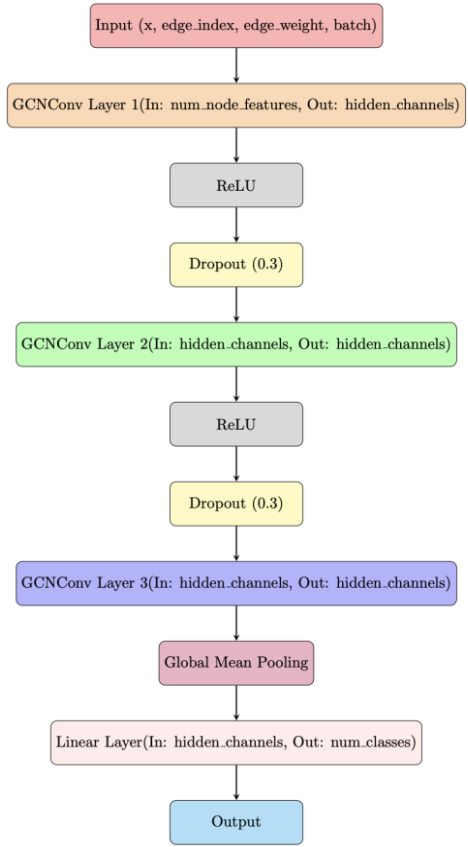 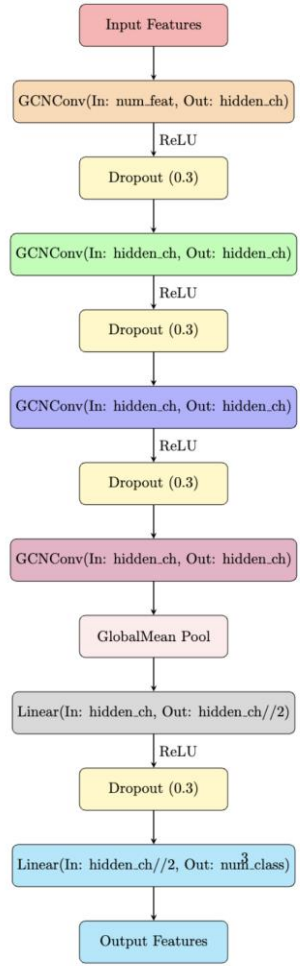

e)                                                            f)

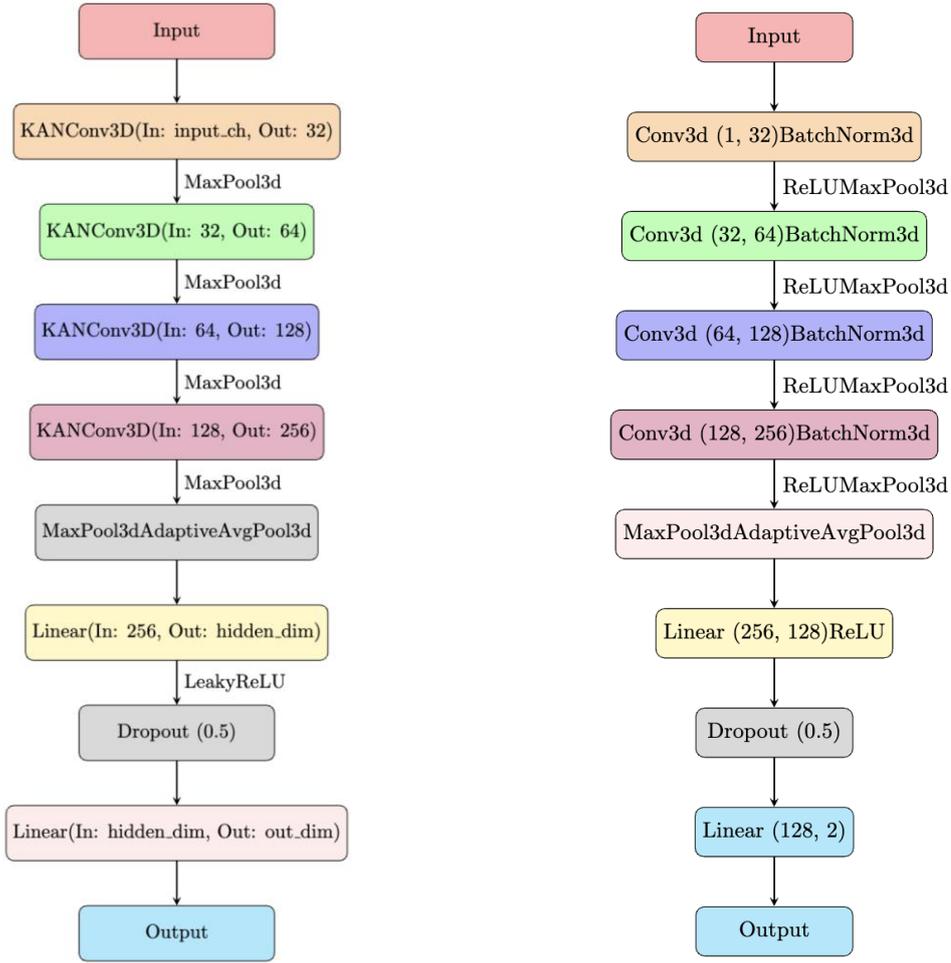

Figure S1: *Architectural diagrams of the 2D and 3D implementations of ConvKAN, CNN, and GCN models.*

Figure S1 illustrates the detailed structures of our six models for PD classification. (a) The 2D ConvKAN presents our 2D implementation, featuring SplineConv2d layers that merge traditional convolutions with learnable non-linear activations. (b) The 2D CNN structure depicts our convolutional approach, highlighting the sequence of 2D convolutional layers, batch normalization, and pooling operations. (c) The 2D GCN diagram illustrates how we process MRI slices as graphs, showing the sequence of GCNConv layers and global pooling. (d) The 3D GCN extends this graph-based approach to volumetric data, demonstrating how entire MRI volumes are analyzed as complex 3D graphs. (e) The 3D ConvKAN diagram showcases our volumetric implementation of the Convolutional Kolmogorov-Arnold Network, featuring KANConv3D layers that combine 3D convolutions with adaptive B-spline activations. (f) Finally, the 3D CNN diagram illustrates the use of standard 3D convolutional operations to process volumetric MRI data. Each diagram provides insights into the layer sequences, input/output dimensions, and key operations specific to each model type and dimensionality, facilitating a comprehensive comparison of our diverse architectural approaches to MRI analysis.

In all models, we employed the Adam optimizer with an initial learning rate of 1e-4 and weight decay of 1e-5. We implemented early stopping with a patience of 15 epochs, monitoring validation loss to prevent overfitting. For 2D models, we used a batch size of 64, while 3D models, due to memory constraints and the Leave-One-Out Cross-Validation strategy, used a batch size of 1.

# 3. Graph Creation for MRI Analysis

Our study employed graph-based approaches for both two-dimensional (2D) and three-dimensional (3D) MRI data analysis. This method is predicated on the hypothesis that structural relationships within brain imagery can be effectively captured and classified using graph representations. By transforming MRI data into graphs, we aimed to incorporate both local and global context, potentially overcoming limitations of traditional convolutional approaches in capturing long-range dependencies and structural relationships.

## 2D Graph Creation

For 2D analysis, we transformed each MRI slice into a graph representation using superpixel segmentation. We extracted 100 axial slices centered on the midbrain from each T1-weighted volume, with each slice resampled to 224x224 pixels using bilinear interpolation.

Superpixel segmentation was employed to generate 1000 superpixels per slice. This technique was chosen for its ability to adhere to image boundaries while maintaining roughly equal-sized segments, which is crucial for creating meaningful graph representations of the MRI data. The algorithm balances spatial proximity and intensity similarity, with parameters tuned to preserve important structural boundaries in brain MRI.

Each superpixel became a node in our graph. For each node, we extracted two key features:

1. Mean intensity: This feature represents the average pixel intensity within the superpixel, capturing the overall brightness or darkness of the region.
2. Relative area: Calculated as the area of the superpixel divided by the total area of the slice, this feature provides information about the size and the importance of each region.

These features were chosen to capture both intensity-based and structural information of the MRI slice, providing a balance between local intensity patterns and global structural characteristics.

Edge connections were established using a k-nearest neighbours approach with k=6, based on the Euclidean distance between superpixel centroids. This choice of k=6 was made to balance local connectivity with some degree of long-range interactions. Edge weights were computed as the inverse of (1 + distance) between connected nodes, encouraging the model to prioritise local connections while still allowing for some long-range interactions.

# 3D Graph Creation

For 3D analysis, we extended the concept of superpixels to volumetric data, using supervoxels instead. Each 3D volume was resampled to a uniform size of 128x128x128 voxels using trilinear interpolation to ensure consistency across different MRI acquisitions and reduce computational complexity while preserving essential structural information.

We generated 1000 supervoxels per MRI volume, balancing the trade-off between capturing fine-grained structural information and maintaining computational efficiency.

For 3D graphs, we extracted three features for each node:

1. Mean intensity: The average voxel intensity within the supervoxel.
2. Relative volume: The volume of the supervoxel relative to the total brain volume.
3. Normalised centroid coordinates: The x, y, and z coordinates of the supervoxel centroid, normalised to the range [0, 1].

These features were chosen to capture not only intensity and size information (as in the 2D case) but also spatial location within the brain volume. The inclusion of centroid coordinates allows the graph to retain information about the spatial arrangement of brain structures, which is crucial for capturing the 3D topology of the brain.

As in the 2D case, edge connections for 3D graphs were established using a k-nearest neighbours approach with k=6, based on the Euclidean distance between supervoxel centroids. The same inverse distance weighting scheme was applied to edge weights.

## Advantages of Graph-Based Representation

The use of superpixels/supervoxels instead of individual pixels/voxels offers several advantages:

1. Computational efficiency: By reducing the number of nodes in the graph, we significantly decrease computational complexity, allowing for faster training and inference.
2. Improved feature representation: Superpixels/supervoxels capture local structural information more effectively than individual pixels/voxels, as they represent coherent regions of the image.
3. Noise reduction: By aggregating features over regions rather than using individual pixel/voxel values, we reduce the impact of noise in the MRI data.

The choice of 1000 superpixels/supervoxels was made empirically to balance detail preservation with computational feasibility. This number allows for sufficiently fine-grained segmentation of brain structures while keeping the graph size manageable for subsequent processing steps.

## Validation of Graph Representations

To ensure our graph representations captured meaningful differences between Parkinson's Disease (PD) and control subjects, we calculated several graph metrics:

1. Average degree: Measures the average number of connections per node.
2. Average clustering coefficient: Quantifies the degree to which nodes tend to cluster together.
3. Average assortativity coefficient: Measures the preference for nodes to attach to similar nodes.

We also quantified the differences between PD and control graphs in terms of:

1. Feature difference: Calculated as the Euclidean distance between the mean node feature vectors of PD and control graphs.
2. Edge count difference: The absolute difference in the number of edges between PD and control graphs, normalised by the total edge count.
3. Node count difference: The absolute difference in the number of nodes between PD and control graphs, normalised by the total node count.

Our analysis revealed substantial distinctions in node attributes between PD and control graphs, with a feature difference of 22.3 for 3D graphs and 18.8 for 2D graphs. The edge count and node count differences suggested structural variations between PD and control brain networks. These metrics confirmed that our graph construction process captured discriminative information for the classification task in both 2D and 3D scenarios.

By carefully considering the balance between local and global information, computational efficiency, and noise reduction, we effectively captured the structural and intensity-based features of brain MRI scans. The validation metrics suggest that these graph representations successfully encode clinically relevant differences between PD and control subjects, providing a solid foundation for subsequent classification tasks.